\documentclass[aps,prb,twocolumn,nofootinbib]{revtex4-1}	
\usepackage{graphicx}
\usepackage{amsmath, amsfonts,amssymb}
\usepackage{color}
\usepackage{ulem}
\newcommand{\be}{\begin{equation}}
\newcommand{\ee}{\end{equation}}

\newcommand{\bea}{\begin{eqnarray}}
\newcommand{\eea}{\end{eqnarray}}

\newcommand{\la}{\langle}
\newcommand{\ra}{\rangle}

\newcommand{\Ket}[1]{|#1\rangle}

\renewcommand{\vec}[1]{{\bf #1}}
\renewcommand{\epsilon}{\varepsilon}

\newcommand{\addFN}[1]{#1}

\begin{document}
\title{Anomalous Floquet Insulators}
\author{Frederik Nathan$^1$, Dmitry Abanin$^2$,  Erez Berg$^3$, Netanel H. Lindner$^4$, Mark S. Rudner$^1$}
\affiliation{$^1$Center for Quantum Devices and Niels Bohr International Academy, Niels Bohr Institute, University of Copenhagen, 2100 Copenhagen, Denmark\\
$^2$Department of Theoretical Physics, University of Geneva, 1211 Geneva, Switzerland\\
$^3$ Department of Physics, University of Chicago, Chicago, IL 60637, USA\\ 
$^4$Physics Department, Technion, 320003 Haifa, Israel}
\date{\today}
\begin{abstract}

Landau's theory of phase transitions provides a framework for describing phases of matter in thermodynamic equilibrium. 
Recently, an intriguing new class of quantum many-body localized (MBL) systems that do not reach thermodynamic equilibrium was discovered. 
The possibility of MBL systems to not heat up under periodic driving, which drastically changes the nature of dynamics in the system, opens the door for new, truly non-equilibrium phases of matter. 
In this paper we find a two-dimensional non-equilibrium topological phase, the anomalous Floquet insulator (AFI), which arises from the combination of periodic driving and MBL. 
Having no counterpart in equilibrium, the AFI is characterized by an MBL bulk, and topologically-protected delocalized (thermalizing) chiral states at its boundaries. 
After establishing the regime of stability of the AFI phase in a simple yet experimentally realistic model, we investigate the interplay between the thermalizing edge and the localized bulk via numerical simulations of an AFI in a geometry with edges. 
We find that non-uniform particle density profiles remain stable in the bulk up to the longest timescales that we can access, while the propagating edge states persist and thermalize.  
These findings open the possibility of observing quantized edge transport in interacting systems at high temperature.

\end{abstract}
\maketitle

At or near equilibrium, the emergence of universal phenomena enables us to organize our description of physical systems in terms of distinct phases of matter.
Intriguingly, a similar phase structure can emerge far from equilibrium, in {\it periodically-driven} quantum many-body systems.
While some of the corresponding ``Floquet phases'' are analogous to phases that occur in equilibrium~\cite{Yao2007, Photovoltaic_graphene, KBRD, Lindner_Nature, Kitagawa2011, Gu2011, Rechtsman2013, Delplace2013, Katan2013, Usaj2014, Kundu2014, Grushin2014, Dehghani2015, Sentef2015, Bukov2016, Claassen2016, Klinovaja2016, EckardtRMP, Wang2013, Jotzu2014, Aidelsburger2015, Flaschner2016}, others, such as discrete time crystals~\cite{Khemani16, Else16, Potter2016, Curt16b, Else2016b, Roy2016, Choi16DTC, MonroeDTC} or the anomalous Floquet-Anderson insulator (AFAI)~\cite{WindingNumber, AFAI, Quelle2017} and its generalizations~\cite{Jiang2011, Kitagawa2012, Hu2015, HarperRoy17,Potter2017,Fidkowski2017,PoBosons17}, display unique dynamical and topological features that cannot occur in equilibrium. 
We label such phases  ``anomalous Floquet phases.''

The fact that stable phases of matter can exist at all in isolated periodically-driven systems is itself a non-trivial statement: in the absence of a heat bath that can extract energy and entropy, such systems are generally expected to continually absorb energy from the driving field and heat towards a featureless infinite-temperature state at long times~\cite{Ponte14,Alessio14,Lazarides14}.
Crucially, in the presence of strong disorder, many-body localization (MBL) may prevent such heating~\cite{Ponte15,Lazarides15,Abanin20161}. 
Despite their localization, MBL systems support a rich variety of symmetry-breaking and topological phases~\cite{HuseSondhi13,BauerNayak}.


Previous works~\cite{Abanin20161, Lazarides15} have shown that MBL may persist in periodically-driven systems when the driving field has a {\it high frequency} and {\it low amplitude}. 
However, the genuinely new phases of Floquet systems (anomalous Floquet phases), {\it cannot} be realized in the high-frequency regime. 
Specifically, anomalous Floquet phases are characterized by nontrivial evolution over the course of a single driving period, which requires the drive frequency to be at most comparable to other energy scales of the system.
In order to realize the full potential of many-body Floquet systems, we thus must understand the conditions under which anomalous Floquet phases  may be realized. 

\begin{figure}
\includegraphics[scale=0.4]{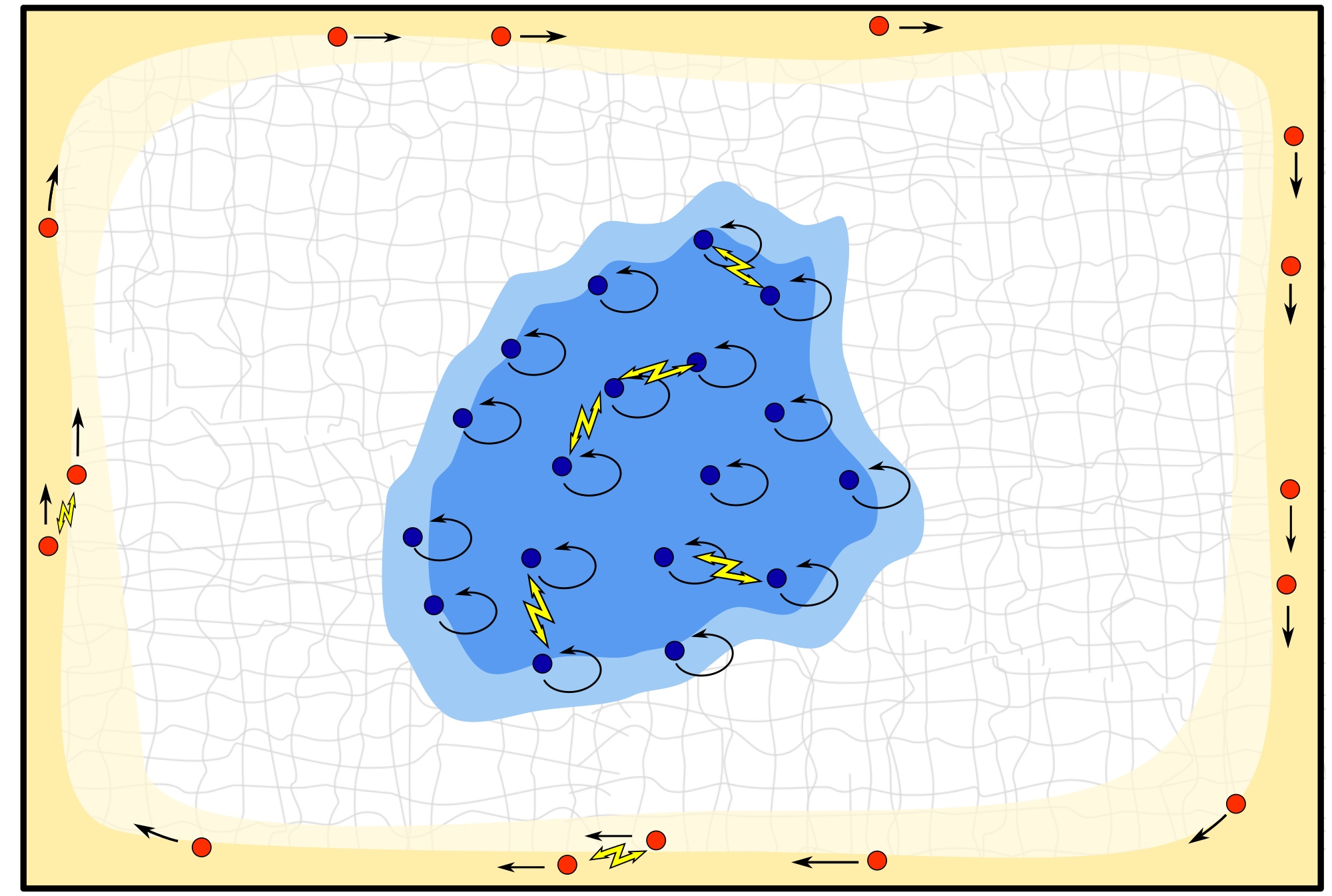}
\caption{
Schematic illustration of the anomalous Floquet insulator (AFI) -- an interacting phase of matter only possible out-of-equilibrium. The bulk states are many-body localized in the presence of disorder and interactions, under conditions discussed in the main text. 
The nontrivial topology of the AFI is manifested in 
chiral edge states that exhibit protected thermalization.
}
\label{fig:Setup}
\end{figure}

In this work we investigate the stability of the two-dimensional (2D) anomalous Floquet insulator (AFI) phase, an interacting version of the AFAI~\cite{AFAI} (see Fig.~\ref{fig:Setup}).  
The AFAI is a topologically nontrivial single-particle anomalous Floquet phase, characterized by a quantized bulk magnetization density~\cite{MagnetizationPaper} 
and protected chiral edge states. 
Here we show that the AFI bulk may be many-body localized in the presence of interactions
\cite{FN:MBLin2D}.
Previously, a variety of  two-dimensional Floquet phases have been studied under the assumption of MBL\cite{Potter2017,Fidkowski2017,PoBosons17}.
Here we directly address the question of whether or not such phases are compatible with MBL. 

To demonstrate MBL, we find conditions under which the original problem 
can be mapped onto an effective high-frequency driving problem in an appropriately constructed rotating frame. 
The same arguments that support MBL in the high-frequency limit~\cite{Abanin20161} then imply MBL of the AFI in the corresponding regime.
This approach can also be applied to establish the stability of  other anomalous Floquet phases, such as discrete time crystals~\cite{Khemani16,Else16} (see  Appendix~\ref{app:TimeCrystals}), and other generalizations of the AFAI~\cite{HarperRoy17,PoBosons17}. 
We support our conclusions with numerical simulations of the long-time dynamics and level statistics of the AFI. 

The crucial differences between the AFI and the AFAI, and some of the AFI's most intriguing properties, are revealed in a geometry with edges.  
First, due to interactions, we expect the topologically protected edge states to give rise to ``protected thermalization'' at the AFI edge, with the particle distribution on the edge rapidly approaching an infinite-temperature-like state. 
 Second, interactions couple thermalizing edge states and localized bulk states, resulting in a non-trivial competition. 
We explore this competition numerically and conclude that, in finite-size samples, the edge and bulk may effectively remain decoupled. 
This opens prospects for realizing quantized edge transport~\cite{AFAI, Kundu} in AFIs at high temperature.


\section{Existence of the anomalous Floquet insulator} 
We first show the existence of the AFI phase for sufficiently weak interactions between particles. 
We consider a system of spinless fermions on a square lattice with two sublattices, $A$ and $B$, described by the following time-periodic Hamiltonian (with driving period $T$):
\be\label{eq:H_general} 
H(t)=H_{\rm id}(t) + H_{\rm dis}(t) + H_{\rm int}, \;\; H(t+T)=H(t). 
\ee
Here $H_{\rm id}(t)$ is the translationally invariant, single-particle Hamiltonian, which realizes the ideal limit of the AFAI~(see Ref.~\onlinecite{WindingNumber} and below).
$H_{\rm dis}$ describes a random on-site disorder potential, which 
stabilizes the AFAI in the absence of interactions~\cite{AFAI}. 
The new ingredient is the two-particle interaction described by $H_{\rm int}$.
 


\begin{figure}
\includegraphics[scale=0.55]{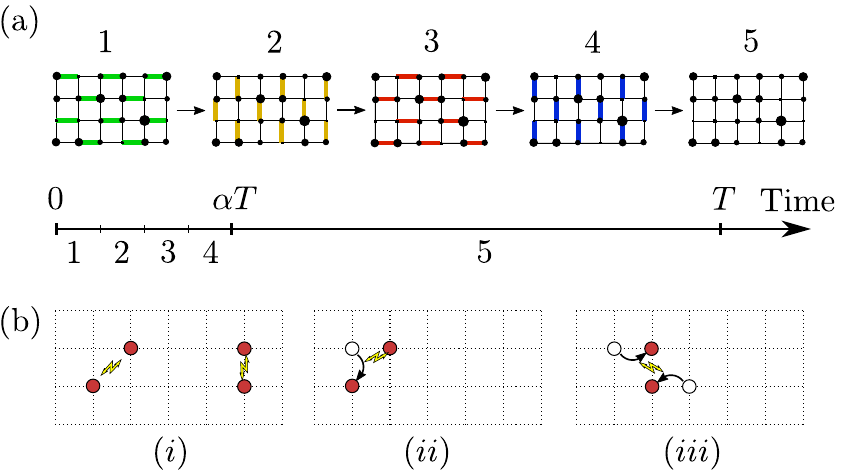}
\caption{
a) Each driving period consists of five segments. 
During the first four segments, time-dependent hopping $H_{\rm id}(t)$ [Eq.~(\ref{eq:Hid})] transfers particles between $A$ and $B$ sublattices, cyclically around plaquettes of the lattice.
Disorder, $H_{\rm dis}$ [Eq.~(\ref{eq:Hdis0})], is applied during the fifth segment, while interactions, $H_{\rm int}$ [Eq.~(\ref{eq:Hint})], are always present. b) Schematic depiction of the terms contained in the transformed interaction Hamiltonian, see Eq.~(\ref{eq:Htransformed}). Three kinds of terms are illustrated: $(i)$ density-density interaction; $(ii)$ hopping with an amplitude dependent on the density of a nearby site; $(iii)$ correlated hopping of pairs of particles. 
}
\label{fig:DrivingProtocol}
\label{fig:HintTerms}
\end{figure}

For concreteness, we consider the following driving protocol, illustrated in Fig.~\ref{fig:DrivingProtocol}a. More general driving schemes will be discussed below.
Each period $T$ is divided into five segments: the first four segments each have duration $\alpha T/4$, and the last segment has duration $(1-\alpha)T$. 
$H_{\rm id}$ acts during the first four segments, while disorder is applied during the last segment; interactions are always present.
Importantly, the parameter $ 0 < \alpha\le 1$ tunes the effective strength of the disorder. 
Below we define how $H(t)$ acts within a single driving period, $0 \le t < T$; its form at later times is obtained from time-periodicity, $H(nT + t) = H(t)$, for any integer $n$.

The Hamiltonian $H_{\rm id}$ consists of hopping terms, which are cyclically applied as illustrated in Fig.~\ref{fig:DrivingProtocol}a:
\be \label{eq:Hid}
H_{\rm id}(t) = J\sum_{ \vec r \in A} \sum_{n=1}^4 f_n(t) (c^\dagger_{\vec r + \vec b_n}c_{\vec r} + {\rm h.c.}),
\ee
where the first sum runs over sites $\vec r$  on sublattice $A$, and $f_n(t)=1$ for $(n-1)\alpha T/4 \leq t < n\alpha T/4$, and $f_n(t)=0$ otherwise. 
The vectors $\{\vec b_n\}$  are given by $\vec b_1=-\vec b_3= (a,0)$  and $\vec b_2 =- \vec b_4 = (0,a)$, where $a$  is the lattice constant. 
The amplitude $J$ is chosen such that the $n$th ``pulse'' perfectly transfers a particle  on site $\vec r\in A$ to site $\vec r+\vec b_n$, and vice versa \addFN{(here and throughout we set $\hbar = 1$)}:
\be\label{eq:J}
J\alpha T/4=\pi/2.
\ee
\addFN{
In this way, $\alpha $ sets the value of the tunneling amplitude: $J=\omega/\alpha $. 
}
We write the disorder Hamiltonian as:
\be\label{eq:Hdis0}
H_{\rm dis}(t)=H_{\rm dis} f_5(t), \;\; H_{\rm dis}=\sum_{{\bf r}} W_{\bf r} c_{\bf r}^\dagger c_{\bf r},
\ee
where $W_{\bf r}\in [-W, W]$ is a random on-site potential, and $f_5(t)=1$ for $\alpha T \le t < T$, and $0$ otherwise.
Finally, we choose $H_{\rm int}$ to consist of nearest-neighbor interactions:
\be\label{eq:Hint}
H_{\rm int}=\lambda\sum_{\langle{\bf r} {\bf r'}\rangle} n_{\bf r}n_{{\bf r'}}.
\ee
In the non-interacting limit, $\lambda=0$, this model is exactly solvable and describes an ideal AFAI with topological edge states and zero localization length in the bulk. 

\subsection{ Rotating frame transformation}
Our goal is to find the conditions when the AFI can be many-body localized. 
Importantly, the driving described above is manifestly {\it not} in the high-frequency limit: condition (\ref{eq:J}) implies that the hopping amplitude $J$ is of the same order as the driving frequency, $\omega=2\pi/T$. 
Therefore, {\it a priori}, the analysis of Ref.~\onlinecite{Abanin20161}  cannot be directly applied. 

We now perform a time-dependent unitary transformation to map our problem onto an equivalent one, which lies in the high-frequency regime as long as $W\ll \omega, J$. 
To avoid complications arising from delocalized edge states, we first consider a system on a torus. 
We transform to a rotating frame in which the fast motion associated with $H_{\rm id}$ is removed:
\be 
\label{eq:Q} |\Phi(t)\rangle = Q^\dagger(t)|\Psi(t)\rangle, \ Q(t)=\mathcal T e^{-i\int_0^t ds H_{\rm id}(s)},
\ee
where $\Ket{\Psi(t)}$ ($\Ket{\Phi(t)}$) is the state in the original (rotating) frame.
We note that 
$Q(T)=I$ is the identity operator: over one full period, evolution with $H_{\rm id}(t)$ alone returns every particle to its initial position. 
It follows that $Q(t)$ is time-periodic: $Q(t)=Q(t+T)$.
    
The time evolution of $|\Phi(t)\rangle$ is generated by a transformed Hamiltonian $\tilde H(t)$, given by $\tilde{H}(t)=Q^\dagger(t) H(t) Q(t)-iQ^\dagger(t)\partial_t Q(t)$. 
By construction, Eq.~(\ref{eq:Q}) gives 
$Q^\dagger H_{\rm id}(t)Q-iQ^\dagger \partial _t Q=0$.
Thus we obtain:
\be \label{eq:Htransformed} 
\tilde{H}(t)=Q^\dagger(t)(H_{\rm dis}(t) + H_{\rm int})Q(t).
\ee
Since $Q(t)$ and $H_{\rm dis}(t)$ are both $T$-periodic, 
$\tilde{H}(t)$  is also time-periodic  with period $T$. 
The periodicity of $Q(t)$ further implies that the Hamiltonian $\tilde{H}(t)$  generates the same Floquet operator as $H(t)$, and therefore the same stroboscopic evolution, $|\Psi(nT)\rangle=|\Phi(nT)\rangle$. 
It  follows that if the system described by $\tilde{H}(t)$  is many-body localized, so is the system described by  $H(t)$. 

With the help of the unitary transformation $Q$, we have eliminated the large-amplitude   term $H_{\rm id}(t)$ from the Hamiltonian. 
The resulting Hamiltonian $\tilde H(t)$  has terms of order $W,\lambda$, which can be much smaller than the driving frequency $\omega$.
In this limit, the system in the rotating frame is in the high-frequency regime, where MBL can be stable with respect to driving. 

\subsection{ Conditions for many-body localization} To establish the conditions for MBL more precisely, we examine the transformed Hamiltonian (\ref{eq:Htransformed}), see also Eqs.~(\ref{eq:Hdis0}) and (\ref{eq:Hint}). Due to the fact that $H_{\rm id}(t)$ acts only during the first four segments of the driving cycle, 
$Q(t)={I}$ 
for all $t\in [\alpha T,T]$. 
Since $H_{\rm dis}(t)$ acts only during the {\it fifth} segment, the disorder Hamiltonian [Eq.~(\ref{eq:Hdis0})] 
is unchanged by the transformation $Q(t)$. 
The disorder term can be decomposed into a time-averaged component $(1-\alpha)H_{\rm dis}$, and a time-dependent component, which changes step-wise at times $t=\alpha T$ and $T$. In the absence of interactions, $H_{\rm dis}(t)$ gives (single-particle) eigenstates that are trivially localized on each site of the lattice.

The transformed interaction Hamiltonian, $\tilde{H}_{\rm int}(t) = Q^\dagger(t)H_{\rm int}Q(t)$, 
has a clear structure including three kinds of terms of extended but finite range (see Fig.~\ref{fig:HintTerms}b):
(i) density-density interactions between nearby sites, (ii) hopping between 
nearby sites with an amplitude that depends on the density on one of the nearby 
 sites, and
(iii) correlated hopping of pairs of particles. 
Explicit expressions for these terms and the ranges over which they act are discussed in Appendix~\ref{app:RotatingFrameTerms}.

Crucially, the transformed interactions remain short-ranged.
All of the terms described above have time-averaged (constant) parts with strengths 
$\sim \mathcal{O}(\alpha \lambda )$, 
  as well as oscillating parts at frequency $\omega$ and higher harmonics, see Appendix~\ref{app:RotatingFrameTerms}.

We proceed in two steps, first analyzing the dynamics generated by the static, time-averaged part of $\tilde{H}(t)$, then investigating the role of the remaining (small) time-dependent terms.
The time-averaged part of $\tilde{H}(t)$ contains 
on-site potential disorder with characteristic scale $W(1-\alpha)$, and one- and two-particle hopping terms induced by interactions, with strength $\sim \lambda\alpha$. 
In the limit $\lambda\alpha\ll W(1-\alpha)$ the delocalizing processes induced by interactions are typically off-resonant, and the (static) system is in the MBL phase~\cite{BAA}. 
At a critical interaction strength $\lambda_c$, the system undergoes a transition into a thermal, delocalized phase.
Thus stability requires:
\be\label{eq:critical}
\frac{\lambda\alpha}{W(1-\alpha)}\leq \kappa_c, 
\ee
where $\kappa_c$ is the critical ratio at which the MBL-delocalization transition occurs. 

As we explain in  Appendix~\ref{app:AlphaLocalization}, the time-dependent terms of $\tilde{H}(t)$ have Fourier components with amplitudes of the order $\alpha \lambda, \alpha W$. 
In the ``high-frequency'' limit, $\omega \gg \alpha  W, \alpha \lambda$, the analysis of Ref.~\onlinecite{Abanin20161} shows that such time-dependent terms do not lead to delocalization. 

The above arguments show that our system exhibits MBL for $ \lambda,  W \ll \frac{1}{\alpha}\omega $, $\lambda < \frac{1-\alpha}{\alpha}W\kappa_c$. 
The AFI thus constitutes a stable anomalous Floquet phase of matter.

\subsection{ Other protocols and phases} The arguments above can be extended to other AFI driving protocols (e.g., if disorder acts throughout the entire driving period, see Appendix~\ref{app:OtherProtocols}). 
The approach we used here can also be used to establish the stability of other anomalous Floquet phases:
in Appendix~\ref{app:TimeCrystals} we apply it to demonstrate the stability of discrete time crystals, which was shown previously by other means~\cite{Khemani16,Else16, Curt16}.

\section{ Numerics: existence of AFI phase}
\label{sec:NumericsLSR}
\begin{figure}
\includegraphics[scale=0.55]{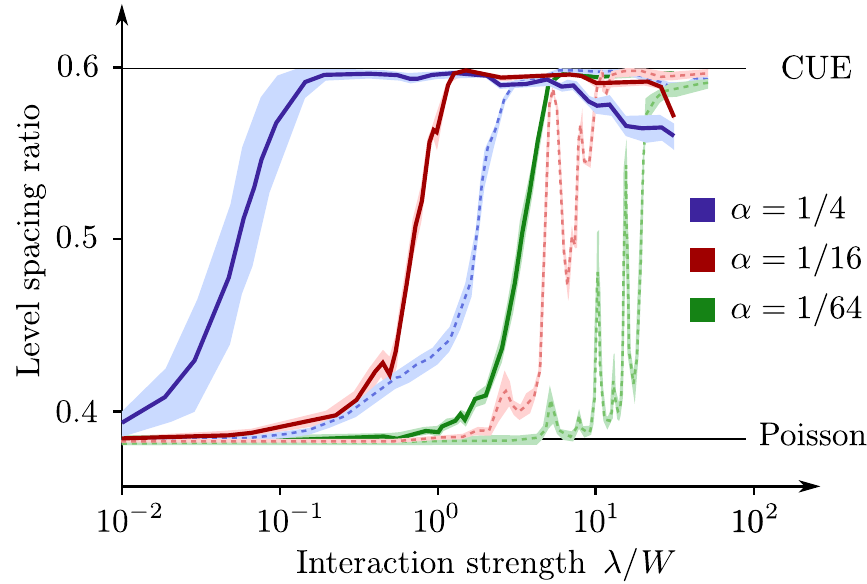}
\caption{
Average level spacing ratio as function of interaction strength, 
in a half-filled system of $4 \times 4$ sites.
Each point on each curve results from averaging over 100 disorder realizations.
Light curves correspond to model (i), with $W=0.1 \omega$, while dark curves correspond to model (ii) with  $W= \omega$.
In all cases, Poisson level statistics, indicating MBL, are observed at low enough interaction strength.
}
\label{fig:LSR}
\end{figure}
 We support the above analytical arguments with numerical simulations. 
To investigate the stability of the phase, we compare two 
driving protocols: 
 (i) the model defined by Eqs.~(\ref{eq:H_general})-(\ref{eq:Hint}), with disorder applied only during the fifth segment,
 and (ii) the same  as (i), but with (constant) disorder  applied throughout the driving cycle.

As an indicator of MBL, we study the quasienergy level statistics of the Floquet operator $U(T)=\mathcal{T}\exp \left( -i\int _0 ^T H(s)\, ds \right)$, obtained via exact evolution~\cite{OganesyanHuse, Ponte15, Alessio14, Lazarides15}. The level spacing ratio around many-body Floquet state $n$ 
is defined as $r_n = \min\{\delta_n/\delta_{n+1}, \delta_{n+1}/\delta _n\}$, where $\delta_n = \varepsilon_{n}-\varepsilon_{n-1}$  is the quasienergy gap below level $n$. 
For a Poisson distribution of levels, this ratio is $\sim 0.4$; for the Circular Unitary Ensemble, it is $\sim 0.6$~\cite{Alessio14}. 


 
We computed the average level spacing ratio by exact diagonalization of the Floquet operator for multiple realizations of the model with 8 particles on a $4\times 4$  square lattice with periodic boundary conditions.
For model (i) we take $W = 0.1 \omega$, and for model (ii) we take $W = \omega$.

Figure~\ref{fig:LSR} shows  the resulting data as a function of $\lambda/W$ (see Appendix~\ref{app:FSSLSR} for the finite-size scaling of the data). 
Each curve shows the mean value of the average level spacing ratio obtained  from an ensemble of $100$ disorder realizations per point~\footnote{For a given value of $\alpha$, the realizations are different for each value of $\lambda$.}, for a fixed value of $\alpha$.
The peaks visible near the transition for $\alpha =\frac{1}{16}$ and $\alpha =\frac{1}{64}$ in model (i) arise due to resonances where the periodic driving  breaks up clusters of $2$, $3$ and $4$ particles that are otherwise bound by the interactions~\footnote{
In  the rotating frame, interactions may bind small droplets of particles together. 
However, when $\lambda = \frac{z\omega}{n}$ for integers $z, n$, the residual periodic driving in the rotating frame can lead to resonances where these clusters break up.
Due to the breaking up of clusters at these values of $\lambda$, the system becomes more delocalized.}.

The data in Fig.~\ref{fig:LSR} show that, for all the values of $\alpha$  we examined, the level spacing ratio converges to $0.38$ for sufficiently small values of $\lambda$, indicative of Poisson level statistics and MBL. 
Additionally, 
the critical value  $\lambda_{\rm critical}/W $ 
at which the localization-delocalization transition occurs shifts upwards for smaller values of $\alpha$, as anticipated above. 
When $\alpha = 1/64$, the system is localized even when the interaction strength is an order of magnitude larger than $W$. 
Smaller values of $\alpha$  will likely push up the transition further.


For a given $\alpha$, the value of $\lambda_{\rm critical}$ in model (ii), where $W = \omega$, is shifted to lower values than in model (i). 
However, $\lambda_{\rm critical}$ remains finite and controllable 
by $\alpha $.
The AFI phase thus appears to extend beyond the regime of the sufficient condition $W\ll \omega$ discussed above. 

\begin{figure}[t!]
\includegraphics[scale=0.6]{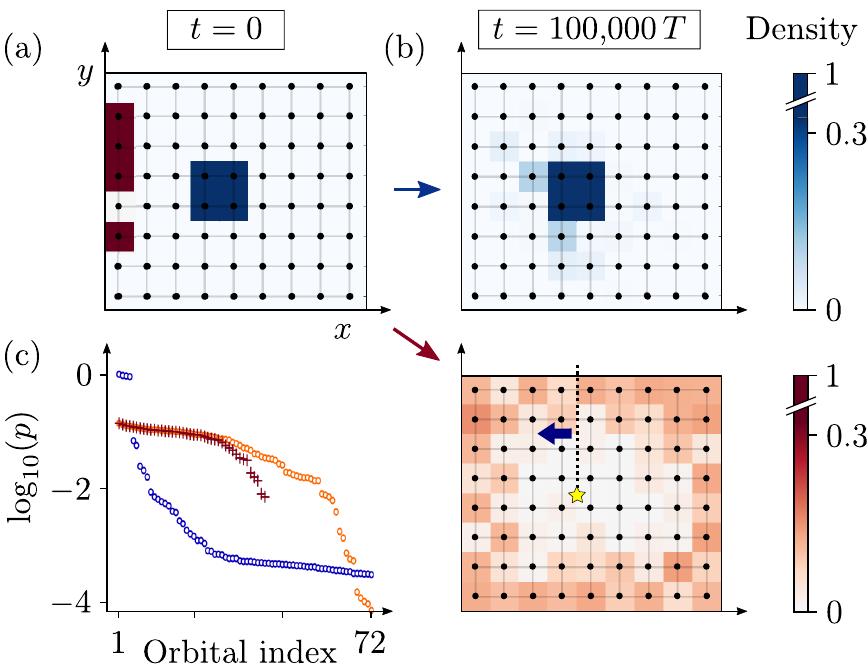}
\caption{ 
Time evolution of 4 particles on a square lattice of $8\times 9$ sites (black dots) with open boundary conditions. 
We simulate model  (ii), with time-independent disorder, and parameter values $W=\omega$, $\lambda = 0.1W$, and $\alpha = \frac{1}{16}$. 
a) The two different initial site occupations  considered,  indicated by red and blue squares. 
b) The cluster initialized in the bulk (blue, upper panel) remains stable over $10^5$ periods.
For the edge initialization (red, lower panel), the particle density is homogenized around the perimeter, with negligible leakage into the bulk. 
\addFN{The dashed line in the lower panel indicates the cut used to calculate the current in Fig.~\ref{fig:Current} (see main text). }
c) Eigenvalues of the one-body reduced density matrix, $\rho^{(1)}_R$.
For the bulk initialization (blue), we take $R$ 
to be the  full lattice; a clear gap between near-unity and smaller eigenvalues indicates localization. 
For the edge initialization, we consider $R$ as 
the full lattice (orange), or only the sites along its edge (red). 
The nearly identical plateaus of eigenvalues in the two cases   indicate thermalization confined to the edge. 
}
\label{fig:TimeEvolution}
\end{figure}

\subsection{ Dynamics of an AFI with edges}  
\label{sec:DynamicsOfAnAFIWithEdges}
So far, we have established the stability of the AFI in a closed geometry without an edge. 
In the non-interacting AFAI in an open geometry (i.e., a geometry with edges), the system's nontrivial topology gives rise to propagating chiral edge states and novel quantized transport phenomena~\cite{AFAI, Kundu}.
Due to the topological and {\it chiral} nature of the edge states, we expect that interactions will lead to  
thermalizing behavior at the edge. 
\addFN{Note that the thermalization of particles on one edge does not preclude a nonzero net current, since the counterpropagating  modes are confined on opposite edges:  even when the particles on one edge thermalize to an effective infinite-temperature state, only one of the edge modes will be populated. 
}
The competition between thermalization on the one-dimensional edge and MBL of the two-dimensional bulk is a subtle and important issue to explore. 
A related problem of an MBL system coupled to a thermalizing edge was recently analyzed in Ref.~\onlinecite{Nandkishore2016}, where the thermal edge was treated as an effective external  bath. 
By comparing the intrinsic time scales of the effective bath  with the energy and time scales of the MBL bulk, the authors of Ref.~\onlinecite{Nandkishore2016} argued that the two-dimensional case supports a phase where the edge thermalizes 
a finite fraction of the system, 
while the remainder of the bulk remains MBL. 
The AFI provides an intrinsic platform for studying this competition.

To gain insight into the dynamics at the edge, we numerically investigated the AFI in an open geometry. 
We simulated model (ii) discussed above, for $4$ particles moving in a rectangle of $9\times 8 $ sites with open boundary conditions~\footnote{For real time dynamics we are able to simulate larger systems than for level statistics due to the smaller number of particles and because full diagonalization is not needed.}.
We initialized the particles either in a droplet of $2\times 2 $ sites in the center of the system, or in sites along the edge (Fig.~\ref{fig:TimeEvolution}a). 
In Fig.~\ref{fig:TimeEvolution}b, we show the corresponding particle densities after time-evolution for $100\, 000$ driving periods. 
Even after this very long evolution the droplet profile has only slightly broadened, indicating that the bulk acts localized on this time scale (and likely indefinitely). 
For the edge initialization, the particle distribution has homogenized around the perimeter~\footnote{In the non-interacting, clean limit, the chiral edge mode resides on alternating sites on the edge. The observed distribution is homogeneous on these sites, with remaining fluctuations on other sites due to finite-size effects.}, and broadened in a narrow strip near the edge.

\begin{figure}
\includegraphics[width = 0.95 \columnwidth]{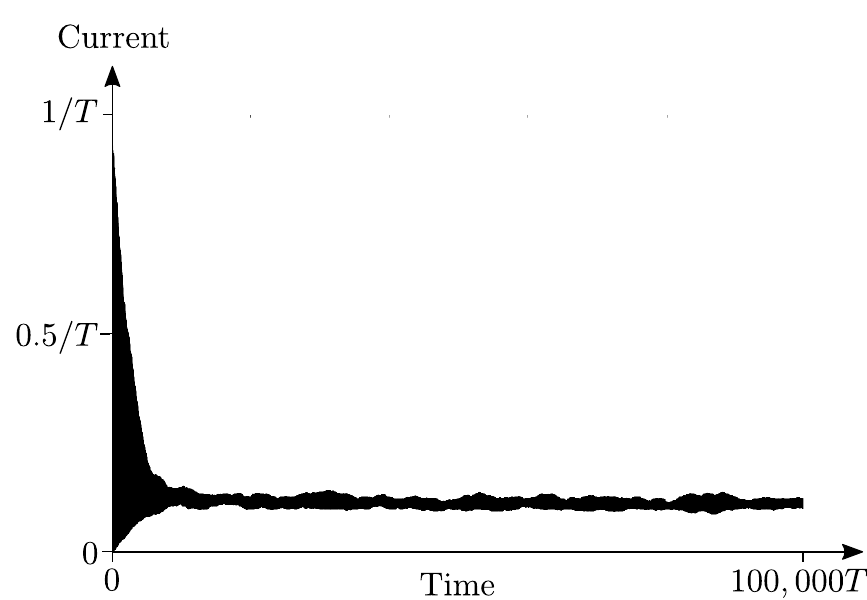}
\caption{
Persistent current in the AFI with particles initialized along its edge.
Here we show the period-averaged current, $I_n$ that flows across the cut in Fig.~\ref{fig:TimeEvolution}d, as a function of time.
}
\label{fig:Current}
\end{figure}

\addFN{
To confirm that the system in Fig.~\ref{fig:TimeEvolution}d carries a nonvanishing circulating current around its perimeter at long times,  we calculate the current flowing across a line that extends from the middle of the system through its boundary (indicated by the dashed line in Fig.~\ref{fig:TimeEvolution}d). 
The period-averaged current, $I_n = \int_{nT}^{(n+1)T} dt I(t)$, where $I(t)$ is the instantaneous current through the cut, is shown in Fig.~\ref{fig:Current}.
The current exhibits large oscillations at short times, due to the fact that the particles that circulate around the perimeter initially have a nonuniform density profile.
At later times, the density of particles along the edge becomes uniform, and the value of the current settles to a nearly constant, nonzero value.
This persistent current is a signature of the chiral nature of the AFI edge.}

To  investigate thermalization at the edge, we first define a region $R$ to be ``thermalized'' if the reduced density matrix on $R$ takes an infinite temperature form, $\rho_R \sim \exp(-\eta\hat{N})$, where $\hat{N}$ is the number operator on $R$ and $\eta$ is a constant that fixes the particle density\addFN{\footnote{\addFN{This form can be obtained by maximizing entropy under the sole constraint of particle number conservation.}}}.
This definition implies, in particular, that on a thermalized region the eigenvalues $p_i$ of the one-body reduced density matrix $\big[\rho^{(1)}_R\big]_{\vec{r}\vec{r}'} \equiv \langle \Psi(t)|c^\dagger_{\vec r} c_{\vec{r}'}|\Psi(t)\rangle$, with $\vec{r},\vec{r}'$ in $R$, are all equal within each particle number sector~\cite{BeraPRL15}: $p_i = N/\mathcal{N}_R$, where $N$ is the number of particles, and $\mathcal{N}_R$ is the number of sites in $R$.

In Fig.~\ref{fig:TimeEvolution}c, we show the eigenvalues of $\rho^{(1)}_R$ after a long time evolution, for both the droplet and edge initializations.
The initial states in both cases are four-particle Slater determinants.
The corresponding one-body density matrices on regions containing all particles would have four unit eigenvalues, with the rest being equal to zero.
For the droplet initialization we choose the region $R$ to be the entire 9$\times$8 lattice; we see that four eigenvalues remain close to one, with only weak correlations among other ``natural orbitals.''
This is a signature of localization~\cite{BeraPRL15}.
For the edge initialization we show the spectra of $\rho^{(1)}_R$ evaluated on a one-site-wide strip running around the perimeter of the system, and on the whole lattice.
For both we find a long plateau of nearly equal eigenvalues signifying thermalization on the edge.
In Appendix~\ref{app:DynamicsNearAnEdge}, we provide additional numerical data for long-time trends in the evolution of the density profile near the edge.

\section{ Discussion}  Our study establishes the 
AFI as a stable anomalous Floquet phase protected by MBL and opens up several directions for future investigations. 
First, our results  \addFN{demonstrate 
 that for finite strengths of  disorder and interactions, the bulk remains localized up to very long times}, even while the edge thermalizes.
This gives promise that the AFI may support quantized transport on all practical/experimental time-scales.
We leave a more detailed 
 investigation of the edge-bulk competition in the thermodynamic limit for a future study.


Second, we found that the chiral AFI's edge hosts {\it protected thermalization}.
The competition between thermalizing and MBL regions is a subject of ongoing debate~\cite{Roeck2017_1}, and the AFI may provide an interesting platform for systematically investigating this interplay.  
For example, consider an AFI punched with holes of circumference $\sim \ell$, typically separated by a distance $L$.
The system can then be viewed as an array of thermalizing regions, each comprised of $\sim \ell$ sites, embedded in a localized background.
Tuning $\ell, L$ allows one to change the volume fraction of thermalizing regions in the system. 
Thus, the {\it geometry} of an AFI sample may be used to control and study thermalization. 


{\it Acknowledgements ---}
 This work was supported by the Swiss National Science Foundation (DA), the Danish National Research Foundation and the Villum Foundation (FN and MR), and the People Programme (Marie Curie Actions) of the European Union's Seventh Framework Programme (No.~FP7/2007-2013) under REA Grant Agreement No.~631696 and the Israeli Center of Research Excellence (I-CORE) ``Circle of Light'' (NL).  NL and EB acknowledge support from the European Research Council (ERC) under the European Union Horizon 2020 Research and Innovation Programme (Grant Agreement No. 639172). MR and EB acknowledge support from CRC 183 of the Deutsche Forschungsgemeinschaft.

\begin{appendix} 
\section{  Hamiltonian in the rotating frame}
\label{app:RotatingFrameTerms}
Here we explicitly compute the transformed interaction Hamiltonian in the rotating frame, $\tilde H_{\rm int}(t)\equiv Q^\dagger(t) H_{\rm int} Q(t)$. 
The interacting part of the Hamiltonian is a sum of terms: 
\be\label{eq:oneterm}
H_{\rm int} = \sum_{\vec{r}, i} H^{(i)}_{{\rm int},\vec{r}}, \quad H^{(i)}_{{\rm int},\vec{r}}=\lambda n_{\vec{r}} n_{\vec{r}+\vec{b}_i},
\ee
where $i = 1, \ldots, 4$, with $\vec{b}_1=-\vec{b}_3=  (a,0)$ and $\vec{b}_2 = -\vec{b}_4= (0,a)$. 
In the rotating frame, the transformed interaction Hamiltonian is computed using Eq.~(\ref{eq:oneterm}) with
\be\label{eq:oneterm2} 
\tilde{H}^{(i)}_{{\rm int},{\vec{r}}}(t)=\lambda \tilde{n}_{\vec{ r}}(t) \tilde{n}_{{\vec{r}}+\vec{b}_i}(t),
\ee
where $\tilde n_{\vec r}(t) \equiv Q^\dagger(t) n_{\vec r} Q(t)$ is the time-evolved site occupation operator. 

We now explicitly compute $\tilde n_{\vec r}(t)$ for the first segment of the driving protocol, $0\leq  t <\alpha T/4$. 
From this we will be able to infer the form of the terms for all later times.
Note that the direction of hopping is opposite for particles initially in the $A$ or $B$ sublattice. 
Therefore, in order to explicitly write $\tilde n_{\vec r}(t)$, we introduce an index $\sigma_{\vec{r}}=1$ for  $\vec r$ in the $A$ sublattice, and $\sigma_{\vec{r}}=-1$ for  $\vec r$  in the $B$ sublattice.
A straightforward computation gives:
\begin{widetext}
\be\label{eq:density_evolved1}
{n}_{\vec r}(t)=\cos^2(Jt) c^\dagger_{\bf r} c_{\bf r}+\sin^2(Jt)c^\dagger_{\vec r+ \sigma_\vec{r} \vec b_1} c_{\vec r + \sigma_\vec{r} \vec b_1}+\frac{i}2 \sin(2Jt)(c^\dagger_{\bf r} c_{\vec r + \sigma_{\vec{r}} \vec b_1}-h.c.), \quad 0\leq t < \alpha T/4.
\ee
\end{widetext}
Note that condition (3) of the main text, $J \alpha T/4 = \pi/2$, yields a simple form for ${n}_{\vec r}(t)$ at the end of the segment: ${n}_{\vec r}(\alpha T/4) = n_{\vec{r} + \sigma_{\vec{r}}\vec{b}_1}$.
Similar expressions are obtained for driving segments 2-4.


The full expression for  $ \tilde  n_{\vec r}(t) \tilde n_{\vec r + \vec b_{i}}(t)$ 
is too cumbersome to write out. 
For the first segment, using Eq.~\eqref{eq:density_evolved1}, it is evident that there are three kinds of terms: 
\begin{itemize}
\item  density-density interaction between nearest and next-nearest neighbor sites. 

\item hopping between nearest-neighbor sites with amplitude that depends on density on one of the nearby sites (terms such as $c^\dagger_{\bf r} c_{\bf r}c^\dagger_{\vec r+\vec b_i} c_{\vec r+\vec b_i-\vec b_1}$),

\item hopping of pairs of particles   (terms such as $c^\dagger_{\bf r} c_{\vec r +\vec b_1}c^\dagger_{\vec r + \vec b_i} c_{\vec r+\vec b_i-\vec b_1}$).

\end{itemize}

In the remaining three segments,  $\tilde  n_{\vec r}(t)$ can be constructed from Eq.~\eqref{eq:density_evolved1}, starting the evolution in each segment with the result of the previous one, 
by $90^\circ$ rotations and translations in the $x$- and/or $y$-directions.
At any time, $\tilde n_{\vec r}(t)$  has its support only on the nearest- and next-nearest neighbor sites of $\vec r$.
In these later segments, the terms in $\tilde H_{\rm int}(t)$  are also of the three types described above, although the distance between coupled sites may be larger than in the first segment.
The   distances between coupled sites in the term $\tilde n_{\vec r}\tilde n_{\vec r+\vec b_i}$  are always bounded by $(1+2\sqrt{2})a$, since $\tilde n_{\vec r}$ has all of its support within a radius of $\sqrt{2}a $ from $\vec r$.  
%

The above discussion shows that  $\tilde H_{\rm int}(t)$ is always  local with a strictly finite range. 
This transformed interaction has an off-diagonal part in the site occupation number basis, whose  
 time-averaged component  has
a magnitude of  order $\alpha\lambda$.
To see this, note that $\tilde H(t)$ only  has off-diagonal components in the interval $0\leq t < \alpha T$, and these have magnitude $\lambda$.

\section{ Other protocols}
\label{app:OtherProtocols}
 The arguments used in this paper can be extended to other driving protocols.   
 As an example, we consider a setup  in which both $H_{\rm dis}$ and $H_{\rm int}$ act throughout the whole driving period. 
We still assume that $W,\lambda\ll \omega$. 

Similar to the analysis above, we employ a unitary transformation $Q(t)$ to eliminate the largest part of the time-dependent Hamiltonian, $H_{\rm id}(t)$. We are left with transformed terms $\tilde H_{\rm dis}(t)$, $\tilde H_{\rm int}(t)$. 
One important difference compared to the main protocol discussed in the text is that the transformed disorder Hamiltonian in this case also contains finite-ranged hopping terms,  of the order $\alpha W$. 
In the absence of interactions ($\lambda=0$), the system is in the localized phase for small disorder $W\ll \omega$, as shown in Ref.~\onlinecite{AFAI}. Moreover, tuning parameter $\alpha$ allows one to tune the localization length in the single-particle problem: at very small $\alpha$ (corresponding to very strong hopping during first four segments of the period), the localization length can be made much shorter than the lattice constant. 

The interaction terms transform in the same way as described in the previous subsection.
Provided $\lambda$ is sufficiently small compared to $W$, these terms will not delocalize the system. 
We note that the presence of single-particle hopping terms originating from the disorder Hamiltonian will reduce the critical value of the interaction strength at which delocalization occurs. 
Residual hopping outside of $H_{\rm ideal}(t)$ (i.e., imperfect hopping ``$\pi$-pulses'') will have a similar effect. 
We thus conclude that AFI phase is generally stable with respect to weak interactions, irrespective of the precise driving protocol.

\section{ Localization controlled by  $\alpha$}
\label{app:AlphaLocalization}
Here we briefly comment on how $\alpha$  controls the localization properties of models (i) and (ii) discussed in the main text.
This analysis applies to both models.
After applying the rotating frame transformation, Eq.~(6) of the main text, 
we write the transformed Hamiltonian $\tilde{H}(t)$, Eq.~(7), as $\tilde H(t) = \bar H + \delta \tilde H(t)$.
Here $\bar H$ is the time-average of $\tilde H(t)$. 
We further decompose $\bar{H}$ as 
$\bar H=H_{\rm int} + \bar{H}_{\rm dis}  + \mathcal O (\alpha W, \alpha \lambda)$, where $\bar{H}_{\rm dis}$ is the time average of $H_{\rm dis}(t)$ over the fifth segment. 
The $\mathcal O (\alpha W, \alpha \lambda)$ corrections arise due to 
the transformation during the window $0 \le t < \alpha T$ where the hopping is applied.

Both $H_{\rm int}$ and $\bar{H}_{\rm dis}$ are diagonal in the site occupation number basis.
The off-diagonal contributions to $\bar{H}$, contained in the $\mathcal O (\alpha W, \alpha \lambda)$ terms, can be made arbitrarily small by taking $\alpha$  small enough. 
In this way we can ensure that, in the absence of the time-dependent terms $\delta \tilde H(t)$, $\bar{H}$ describes a many-body localized system.

Next, we consider the oscillating part of $\tilde{H}(t)$, $\delta \tilde H(t)$, which
has a magnitude of order $W,\lambda$, varies rapidly in the interval $0 \le t < \alpha T$, and is constant for the rest of the period.  
Turning to the Fourier transform of $\delta \tilde H(t)$, these properties dictate that its  $n$-th Fourier component 
is of order $\alpha W, \alpha \lambda$  
for $|n|\lesssim \frac{2\pi }{\alpha}$, and falls off as $1/n$ for large $n$. 
In the limit $\omega 
 \gg \alpha \lambda, \alpha W$, 
 even the lowest harmonics correspond to high frequencies in the rotating frame, and therefore the system remains localized.
For $\omega$ comparable to or greater than $W, \lambda$, the amplitude of the 
 oscillating terms can be made arbitrarily small by taking $\alpha \rightarrow 0$.
This again brings the system into the Floquet-MBL regime.

\section{Stability of time crystals}
\label{app:TimeCrystals}
 To demonstrate the universality of our approach, we now outline an argument for the stability of the discrete time crystal (DTC)~\cite{Khemani16,Else16}.
The  DTC is an example of an anomalous Floquet phase where the discrete time-translational symmetry of the drive, $t\to t+T$, is broken. 
We note that the stability of DTCs has been previously investigated numerically and through other analytical arguments in Refs.~\cite{Khemani16,Else16,Curt16}. 

First, following Ref.~\onlinecite{Else16}, we consider a solvable driving protocol for a one dimensional spin-$1/2$ chain, which illustrates the basic physics of the DTC: 
\be\label{eq:DTC_solvable}
H_0(t)=f(t) H_{x}+[1-f(t)]H_{\rm dis},
\ee 
where $f(t)=1$ for $t\in [nT,nT+T/2]$ and zero otherwise.
With this protocol, the first (second) term in the Hamiltonian is turned on during the first (second) half-period. 
The Hamiltonian $H_{x}$ induces a global spin rotation around the $x$ axis.
The strength of the uniform applied $x$-field is chosen such that the evolution over the first half-period gives a perfect $\pi$-pulse:
 \be\label{eq:flip}
 H_{x}=\frac{\pi}{T} \sum_i \sigma_i^x.
 \ee
 The disorder Hamiltonian is chosen as a random, nearest-neighbor Ising interaction:
 \be\label{eq:Ising}
 H_{\rm dis}=\sum_{\la ij \ra} J_{ij} \sigma_i^z \sigma_j^z, \ \ J_{ij}\in [\bar{J}-W,\bar{J} + W],
 \ee 
where $\bar{J}$ sets the average interaction strength, and $W$ is the width of the distribution of random couplings.

The evolution generated by protocol (\ref{eq:DTC_solvable}) can be solved exactly.
For simplicity, consider an initial product state $|\Psi(0)\ra=\bigotimes_i|\sigma_i \ra\equiv |\{\sigma_i\}\rangle$, in which each spin points up or down along $z$, $\sigma_i=\pm 1$. 
(The argument works for all such configurations.)
During the first half-period, each spin is flipped: $|\{\sigma_i \}\ra\to |\{-\sigma_i \}\ra$.
Note that the state remains a product state in the $z$-basis.
During the second half of the period, the state acquires a dynamical phase due to the Ising interaction (\ref{eq:Ising}). 
Over the next driving period, a second $\pi$-pulse flips all spins back to their initial configuration.
In total, the local $z$-projection $\la \sigma_i^z \ra$ of each spin oscillates with twice the period of the drive. 
Remarkably, this behavior is stable with respect to generic $T$-periodic perturbations of the Hamiltonian. 

To show the stability of DTCs using our approach, we add a small local, but otherwise generic perturbation to the time-dependent Hamiltonian (\ref{eq:DTC_solvable}):
\be\label{eq:perturbed}
H(t)=H_0(t)+\lambda H_{\rm pert}(t), \,\,\, \lambda\ll 1.  
\ee
We assume that $H_{\rm pert}(t)$ shares the same periodicity as the drive, $H_{\rm pert}(t+T) = H_{\rm pert}(t)$. 

Similar to the AFI discussed in the main text, this problem is not in the high-frequency limit. 
More specifically, the frequency $\omega$ is comparable to the amplitude of the local field in $H_{x}$, as it must be in order to induce a spin flip during one half-cycle. 
Similar to our analysis of the AFI, we move to a rotating frame which removes the large-scale micromotion (i.e., the repeated $\pi$-pulses).
This is accomplished via the transformation $|\Phi(t)\ra=S^\dagger(t)|\Psi(t)\ra$, with
\be\label{eq:S}
S(t)=\mathcal T e^{-i\int_0^t ds f(t) H_{x}(s)}. 
\ee
We note that $S(nT)=P^{n\, (\rm mod\ 2)}$, where 
\be\label{eq:p-operator}
P=\prod_i (i\sigma_i^x)
\ee
is a global spin-flip operator.

Taking into account the fact that the Ising disorder Hamiltonian commutes with $S(t)$, the Hamiltonian in the rotating frame is given by:
\be\label{eq:Heff}
\tilde H(t)=[1-f(t)]H_{\rm dis}+S^\dagger(t)H_{\rm pert}(t)  S(t). 
\ee
Interestingly, the  periodicity of the dressed perturbation $\tilde{H}_{\rm pert}(t) = S^\dagger(t)H_{\rm pert}(t)  S(t)$ may be reduced to $2T$-periodicity. 
This is easy to see, for example, for $H_{\rm pert}(t)=g(t)\sum_i \sigma_i^y$, using Eq.~(\ref{eq:p-operator}) and $g(t + T) = g(t)$. 
Importantly, this term remains {\it local}, since $S(t)$ [Eq.~(\ref{eq:S})] simply describes spin rotations over the first half-period.

Having eliminated the large term (\ref{eq:flip}), we see that for sufficiently small interactions, $\bar{J}, W\ll \omega$, the transformed Hamiltonian in the rotating frame {\it is} in the high-frequency driving regime. 
Therefore, by the perturbation theory of Ref.~\onlinecite{Abanin20161}, we can argue that for a sufficiently weak perturbation, $\lambda\ll W$,  the system is in the MBL phase. 
Thus the time-evolved wave function (in the rotating frame), $|\Phi(t)\ra=\tilde U |\Phi(0)\ra$, with $\tilde U(t)={\mathcal T}e^{-i\int_0^t \tilde H(s)ds}$, retains the memory of the initial state. 

Finally, we discuss why MBL of the transformed problem (\ref{eq:Heff}) implies persistent oscillations of physical observables with a doubled period. 
As above, choose the initial state to be a product state $|\Psi(0)\ra=  |\{\sigma_i\}\rangle$. 
Then, MBL implies that the local magnetization evaluated in the rotating frame, $\la \tilde\sigma_i^z(t)\ra :=\la \Phi(t)| \sigma_i^z |\Phi(t)\ra$, remains close to its initial value for all $t\to \infty$ (at least in the strong-disorder limit $\lambda\ll W \ll \omega$). 
 Then, using Eq.~(\ref{eq:p-operator}) and the fact that $P^\dagger \sigma_i^z P=-\sigma_i^z$, we relate the physical local magnetization at stroboscopic times to $\la \tilde\sigma_i^z (t)\ra$:
\be
\la\sigma_i^z(nT) \ra=\la \Psi(nT)| \sigma_i^z|\Psi(nT) \ra=(-1)^n  \la \tilde\sigma_i^z (nT)\ra.\vspace{0.1 in}
\ee
Since $\la \tilde\sigma_i^z (nT)\ra$ remains close to its initial value, we have shown that the magnetization oscillates with period $2T$, persisting to the limit $t\to \infty$.


\section{Dynamics near an edge}
\label{app:DynamicsNearAnEdge}
\begin{figure}
\includegraphics[width=\columnwidth]{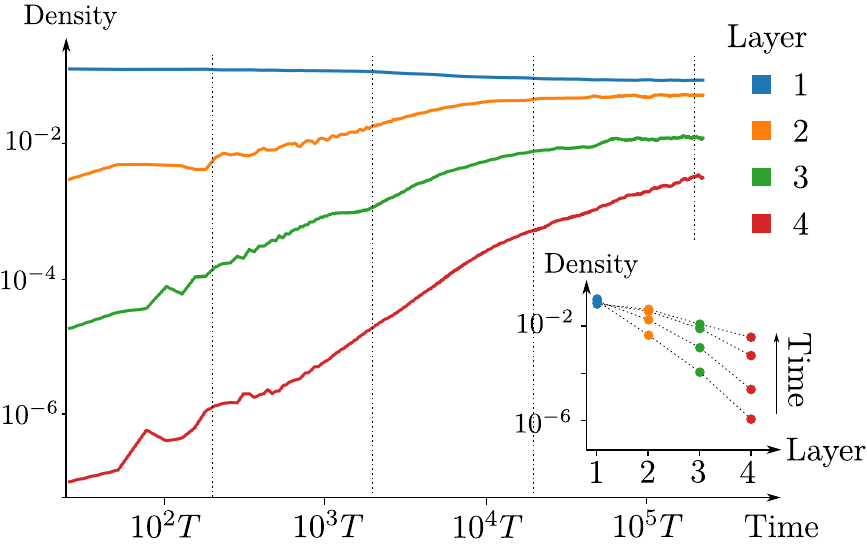}
\caption{
Average density as a function of time in the 4 concentric layers of the lattice, for the simulation depicted in Fig.~\ref{fig:TimeEvolution} (see main text for  further details).
Inset: Density  as a function of layer index at  the times indicated by vertical dashed lines in main panel. }
\label{fig:DensVsLayer}
\end{figure}

In this appendix, we explore the long-time trends in the evolution of the density profile in the model studied in Sec.~\ref{sec:DynamicsOfAnAFIWithEdges}, for the initialization where the particles were located on the edge (red in Fig.~\ref{fig:TimeEvolution}a). 
Specifically, from the time-evolution of the system,  we  extracted the average density in concentric layers of  the  lattice as a function of time. 
We divided the $9\times 8$ rectangular lattice into $4$ layers, with  layer $1$ containing the  sites on the lattice's edge ($30$ sites in total), layer $2$ containing the sites located one lattice constant from the edge ($22$ in total), layer $3$ containing sites located $2$ lattice constants from the edge ($14$ in total), and layer $4$ containing the 6 innermost sites.
In Fig.~\ref{fig:DensVsLayer}, we plot the average density in each layer, as a function of time, for the first $200,000 $ driving periods.
As can be seen, the density in each of  the inner  layers appears to grow as  a power law,  with different exponents for the distinct layers, until after approximately $10,000 $ driving periods.
After this point the average densities in the layers begin to saturate.  
From the system sizes we have accessed, we can not confirm whether the saturation is intrinsic or due to finite size effects. 
To highlight the density's dependence on layer index, in the inset we show the density   versus layer index at four different times (indicated by vertical dashed lines in main panel). 

\section{Finite-size scaling of level spacing ratio}
\label{app:FSSLSR}

\begin{figure}
\includegraphics[width=\columnwidth]{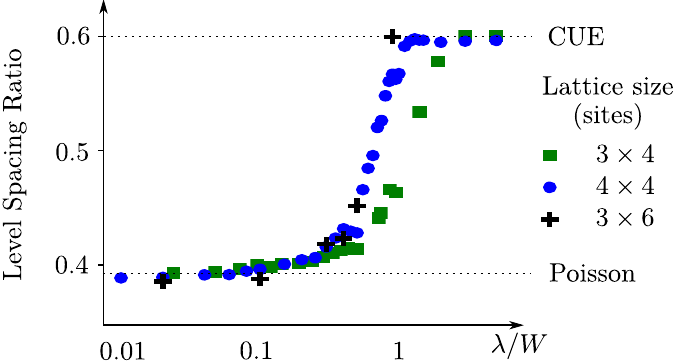}
\caption{Average quasienergy level spacing ratio, as function of interacting strength $\lambda$, for the model in Sec.~\ref{sec:NumericsLSR}, at $3$ different choices of lattice size. 
The parameters $W$ and $\alpha $ were set to $W=\omega/15$, and $\alpha =1/16$. 
See main text for further details. 
}
\label{fig:FiniteSizeLSR}
\end{figure}

Here we investigate the  finite-size scaling of  the quasienergy level spacing ratio for the model studied in Sec.~\ref{sec:NumericsLSR} (see also Fig.~\ref{fig:LSR}). 
Specifically, here we provide data for the model on  half-filled lattices with $3\times 4$, $4\times 4$  sites and $3 \times 6$ sites. 
The parameters for the model  were set to  $W= \omega/15$, $\alpha = 1/16$, while the interaction strength $\lambda$ was varied.

In Fig.~\ref{fig:FiniteSizeLSR} we show the average level spacing ratio, as a function of $\lambda$,  for the $3$ lattice sizes mentioned above.
For the $3\times 4$ ($4\times 4$) system, each data point 
 was obtained from the average level spacing ratio over the full spectrum for 20 (4) disorder realizations. 
For the $3\times 6$ system, 
each data point was computed from the average level spacing ratio of $3000$ adjacent levels in the quasienergy spectrum  for a single disorder realization at the given value of $\lambda$. 

As can be seen in Fig.~\ref{fig:FiniteSizeLSR},  with decreasing interaction strength, there is a clear crossover of the level spacing ratio from the value corresponding to the Wigner-Dyson circular unitary ensemble (CUE) to Poisson statistics. 
This behavior is indicative of a delocalization-localization transition. 
We also note that the transition appears to sharpen with increasing system size. 
Finally, although the data for the $3\times 6$ system are sparse, the data in Fig.~\ref{fig:FiniteSizeLSR} suggest that  the level spacing for the $3$ series cross near $\lambda \approx 0.2 W$. 
However, due to the limited data, this observation is not  conclusive. 
In particular, the crossover may be due to the varying aspect ratio of the system over the 
$3$  system sizes we probed.
In order to establish the existence of  a localization-delocalization transition in the thermodynamic limit, a more extensive numerical study is thus required.

\end{appendix}

\end{document}